\documentclass[useAMS,usenatbib]{mn2e}

\title[Gemini Near Infrared Integral Field Spectroscopy of the NLR of ESO\,428-G14]{Gemini Near Infrared Integral Field Spectroscopy of the Narrow-Line Region of ESO\,428-G14: kinematics, excitation and the role of the radio jet}

\author[Rogemar A. Riffel et al.]
{Rogemar A. Riffel$^{1}$\thanks{E-mail: rogemar@ufrgs.br (RAR);
thaisa@ufrgs.br (TSB); cwinge@gemini.edu (CW); faustokb@if.ufrgs.br
(FKBB)}, Thaisa Storchi-Bergmann$^{1}$\footnotemark[1]\thanks{Visiting
Astronomer, Cerro Tololo Inter-American Observatory, National Optical
Astronomy Observatories, which are operated by AURA, Inc., under a
cooperative agreement with the National Science Foundation.},
Cl\'audia Winge$^{2}$\footnotemark[1] and \newauthor
 Fausto K. B. Barbosa$^{1}$\footnotemark[1]\\ 
$^{1}$Universidade Federal do Rio Grande do Sul, IF, CP 15051, Porto
Alegre 91501-970, RS, Brazil.\\ 
$^{2}$Gemini Observatory, c/o AURA Inc., Casilla 603, La Serena, Chile.}

\usepackage{graphicx}

\begin{document}

\date{Accepted 1988 December 15. Received 1988 December 14; in original form 1988 October 11}

\pagerange{\pageref{firstpage}--\pageref{lastpage}} \pubyear{2002}

\maketitle

\label{firstpage}

\begin{abstract}

We present two-dimensional (2D) gas kinematics and excitation of the  inner 300\,pc of the Seyfert galaxy ESO\,428-G14 at a sampling of  14\,pc$^2$, from near-infrared spectroscopic observations at  R$\approx$6000 obtained with the Integral Field Unit of the
Gemini Near-Infrared Spectrograph. From measurements of fluxes and  profiles of the emission lines [Fe\,{\sc ii}]$\lambda 1.257 \mu$m,  Pa$\beta$, H$_2\lambda 2.121 \mu$m and Br$\gamma$, we construct
2D maps of line intensities and ratios, radial velocities and velocity
dispersions.  Emission lines ``tomography" is provided by velocity slices obtained across the line profiles, a unique capability of IFUs, which allows the mapping of not only of peak velocities but  including also the wings.
We compare these maps with a previously published high
spatial resolution radio map and find a tight relation between the  radio structure and the emission-line flux distributions and  kinematics, revealing that the radio-jet plays a fundamental role not  only in shaping the NLR but also in the imprint of its kinematics.
Blueshifts of up to 400\,km\,s$^{-1}$ and velocity dispersions of up  to 150\,km\,s$^{-1}$ are observed in association with the radio jet at  position angle PA\,=\,129$^\circ$, which is also the PA of the  photometric major axis of the galaxy.  We  conclude that the radio jet is launched at a small angle relative to  the galactic plane, with the NW side slightly oriented toward us. This  angle is small enough for the radio jet to shock and compress the gas  in the plane of the galaxy, and for the nuclear continuum to ionize and heat it. 
The distinct kinematics and flux distributions observed for the different emission lines suggest different origins for their emission. The [Fe\,{\sc ii}] shows the largest  blueshifts and velocity dispersions and its flux distribution is concentrated along the jet, while the H$_2$ shows the lowest velocity dispersions and has additional flux contribution from regions beyond the jet.  Both X-rays emitted by the active galactic nucleus and shocks produced by the radio jet can excite the H$_2$ and [Fe\,{\sc ii}] emission lines.
 We use the 2D  velocity dispersion maps to estimate upper limits to the contribution of the radio jet to the excitation of [Fe\,{\sc ii}] and H$_2$ which may reach 90\% for [Fe\,{\sc ii}] and 80\% for H$_2$ in the jet region. The [Fe\,{\sc ii}]/Pa$\beta$ emission-line ratios and the association of the [Fe\,{\sc ii}] flux 
distribution and kinematics with the radio structure supports a stronger contribution of the radio  jet to the [Fe\,{\sc ii}] excitation than to that of H$_2$.  In the regions beyond the jet the observations favor X-ray excitation.

\end{abstract}


\section{Introduction}

The Narrow-Line Region (NLR) of Seyfert galaxies is one of the best
probes of the mechanisms in operation in the surrounding of accreting
supermassive black holes in galaxies. The excitation and dynamics of
the inner NLR gas can reveal how radiation and mass outflows from the
nucleus interact with circumnuclear gas. Until recently, the
measurement of these properties had to rely on either narrow-band images
\citep[e.g.][]{wilson93, capetti96, schmitt96} or long-slit
spectroscopy. Long-slit spectroscopic studies provide information on
the origin and excitation of the emission lines \citep[e.g.][]{veilleux97,
rodriguez-ardila05} as well as on the gas kinematics \citep[e.g.][]{
winge97,winge99,crenshaw00,hutchings98,
kaiser00,nelson00}, but are restricted to the locations covered
by the long-slit. 

\citet{mulchaey96} and \citet{ferruit00} have shown that obscuration
can affect the optical morphology of the emitting gas region, a
problem that can be alleviated by using infrared lines to map the
NLR. Relevant emission-lines in the near-IR include [Fe\,{\sc
ii}]$\lambda$1.257$\mu$m and 1.644$\mu$m, HI lines such as
Pa\,$\beta$ and Br\,$\gamma$ and molecular hydrogen lines
such as H$_2\lambda$1.957$\mu$m and H$_2\lambda$2.121$\mu$m, which can be used to map the gas kinematics and excitation \citep{storchi-bergmann99,winge00}.

Studies based on the above lines \citep{forbes93,blietz94} have revealed a correlation between the [Fe\,{\sc ii}] and the radio emission,
indicating that shock excitation by the radio jets is a likely
mechanism for production of the [Fe\,{\sc ii}] emission in these
objects, although other works have favoured photoionization by the nuclear source
\citep{simpson96,das05,das06}. H$_2$ emission is also strong in
AGNs \citep[e.g.][]{veilleux97,rodriguez-ardila04,rodriguez-ardila05}
and can be used to probe the distribution of warm molecular gas.  To further progress in this area and investigate the origin of the near-IR emission lines it is necessary to spatially resolve the kinematics and excitation of the emitting gas.

The multiplexing capability necessary to map, simultaneously, the
gaseous distribution, excitation and kinematics at such small scales
with sufficient spectral resolution is now provided by the
integral field units (IFU) in operation in large telescopes. In
this work we use the IFU of the Gemini Near-Infrared Spectrograph
which provides two dimensional (hereafter 2D) mapping in 
the near-IR (thus minimizing the effect of reddening) to study the circumnuclear emitting gas 
 of the Seyfert galaxy ESO\,428-G14. We present
2D flux, radial velocity and velocity dispersion maps in
four strong emission lines:
[Fe\,{\sc ii}]$\lambda 1.257 \mu$m, Pa$\beta\lambda 1.282\mu$m, H$_2
\lambda 2.121 \mu$m and Br$\gamma\lambda 2.166\mu$m. The R$\approx$6000 spectral resolution has allowed us also to obtain velocity slices across the emission-line profiles in order to better probe the velocity fields. 

We present in addition  broad-band {\it J} and {\it K} images and a
narrow-band image in the line [Fe\,{\sc ii}]$\lambda 1.257 \mu$m
obtained with the CTIO Blanco Telescope. We explore the relation
between reconstructed images in the near-IR emission lines, their
kinematic maps, the narrow-band [Fe\,{\sc ii}] image and previously
published optical [O\,{\sc iii}]$\lambda 5007$ 
narrow-band and radio images, in order to investigate the excitation
mechanism of the near-IR lines, in particular [Fe\,{\sc ii}] and
H$_2$. 

ESO\,428-G14 is an S0 galaxy which was classified as Seyfert 2 by \citet{bergvall}. High resolution VLA maps show predominantly a two-sided and asymmetric radio jet oriented approximately along the
galaxy line of nodes at position angle PA$\approx\,129^{\circ}$, in which the NW side is less extended and terminates in a bright hot spot at 0$\farcs$75 from the nucleus, while to the SE the radio emission is fainter, the jet bends to N and its extent is at least three times longer   \citep{ulvestad,falcke96,falcke98}. [O\,{\sc iii}] and H$\alpha$+[N\,{\sc ii}] images obtained with the Hubble Space Telescope show extended emission well aligned with the radio jet with stronger emission to the NW as observed in the radio. The ratio  [O\,{\sc iii}]/(H$\alpha$+[N\,{\sc ii}] ) shows a bipolar structure with larger values to the SE \citep{falcke96,falcke98}.
We adopt  a distance for this galaxy of 19\,Mpc \citep{falcke96} for which 1$^{\prime\prime}$ corresponds to 92\,pc at the galaxy. 

This paper is organized as follows: in Section 2 we describe the
observations and data reduction. In Section 3 we present
the near-IR spectra and images, a comparison with previous optical
narrow-band and radio images and the kinematics of the gas. In Section 4, we discuss the results
and their implications for the origin of the [Fe\,{\sc ii}] and  H$_2$
emission and in Section 5 we present the conclusions of this work. 

\section[]{Observations and Reductions}

\subsection{Integral Field Spectroscopy}

The spectroscopic data were obtained with the Gemini Near Infra-Red
Spectrograph (GNIRS) \citep{elias98} Integral Field Unit (IFU) in
December 2004 under the instrument science verification program
GS-2004B-SV-26,  and comprise two sets of observations centred at
$\lambda$=2.21$\mu$m and 1.27$\mu$m. We have used the 111 l/mm grating
with the Short Blue Camera (0$\farcs$15/pixel) which gives a
resolving power of R=5900. The GNIRS IFU has a rectangular field of
view, of approximately 3$\farcs$2\,$\times$\,4$\farcs$8, divided into 21
slices. At the detector, the slices are divided along their length
into 0$\farcs$15 square IFU elements.  For our observations, the major
axis of the IFU was oriented along the PA $\approx129^{\circ}$, which is
the orientation of the radio jet. The observing procedure followed the
standard Object--Sky--Sky--Object dither sequence, with off-source sky positions since
the target is extended, and individual exposure times of
600s. One of the IFU slicing mirrors was damaged during assembly and
presents a lower (20\% of the nominal value) transmission region, about
1$\farcs$8 in length, offset 0$\farcs$5 from the center, therefore
small spatial offsets in both directions were added between exposures
to ``fill in'' the signal in this region. Telluric standard stars were
observed immediately after the target. The basic observing
information is shown in Table \ref{obs}. Conditions during the
observations were good for the night of Dec. 27 (clear skies, image
quality (IQ) $\sim$0$\farcs$60), and patchy clouds (but stable guide counts)
with similar IQ for the Dec. 28 dataset. 

\begin{figure}
\centering
\includegraphics[angle=-90, scale=0.55]{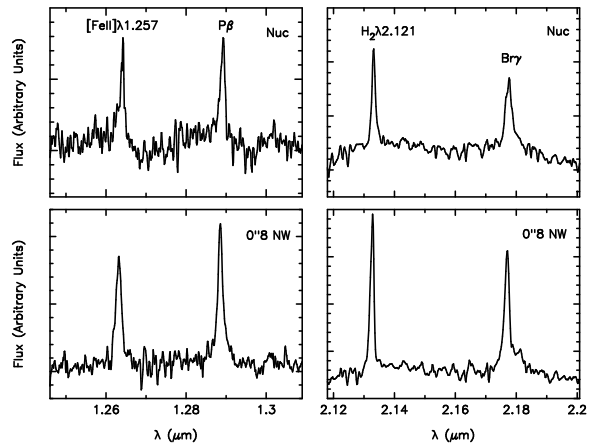}
\caption{Top: Nuclear spectra for ESO\,428-G14 in the {\it J}-band (left)
and  {\it K}-band (right). Bottom: Spectra at 0$\farcs$8\,NW from the
nucleus, which is the position of the radio continuum emission peak.}

\label{espectro}  
\end{figure}

The data reduction was accomplished using tasks contained in GNIRS
package which is part of the GEMINI IRAF package as well as generic
IRAF tasks. Through the reduction tasks we have performed trimming,
flat-fielding, sky subtraction, wavelength and s-distortion
calibrations. We have also removed the telluric bands and
flux calibrated the frames using the star HR4023 (A2V spectral type)
as a ``relative" flux standard. 
The cosmic ray cleaning was done before the sky
subtraction using  the algorithm  described by \citet{vandokkum}.  In Fig.\,\ref{espectro}
we present the nuclear spectra   and the spectra at 0$\farcs$8\,NW from the nucleus, which is the
position of the peak of the radio continuum emission within an aperture of 0$\farcs$45 diameter.

\begin{table}
\caption{Observations}
\centering
\begin{tabular}{c c c c c}
\hline
Instrument/& Date &         & Total Exp.&  \\
Telescope  & (UT) &  Filter & time (s)   \\
\hline
GNIRS/Gemini  & 2004/12/27 & K$_-$G0503 & 2400  \\
GNIRS/Gemini  & 2004/12/28 & J$_-$G0505 & 1200  \\
OSIRIS/Blanco & 2000/02/15 &  K         &   80  \\
              &            &  J         &   80   \\
              &            & 1.06       &  270 \\
              &            &[Fe\,{\sc ii}] 1.257      & 1800  \\
\hline
\end{tabular}
\label{obs}
\end{table}

\subsection {Imaging}

Broad-band {\it J} and {\it K} images and narrow band images centered
on the emission line [Fe\,{\sc ii}]$\lambda1.257$ and adjacent
continuum were obtained in February 2000 at the Cerro Tololo
Interamerican Observatory (CTIO) using the 4-m Blanco telescope.  The
observations were performed with the infrared imager/spectrometer
OSIRIS using a 1024\,$\times$\,1024 HgCdTe array at an angular scale
of 0\farcs16\,pix$^{-1}$. Details of the observations are presented in
Table \ref{obs}. The angular resolution was 0\farcs8-0\farcs9 (FWHM of
standard star images). 

The observing procedure consisted of taking a sequence of 4--9
dithered images on source with sky images taken before
or after the object sequence.  Individual sky and object images were first divided by the
flat-field. We then median combined the  sky images and subtracted
them from the corresponding object images. The individual images in
the same filter were then aligned using as many common point
sources (stars and the nucleus) as  possible and finally averaged. We
have used the task {\tt imcoadd} from the GEMINI package to  perform
this alignment/averaging procedure. 

Photometric calibration was performed for both broad and narrow band
filters using standard stars. Correction for extinction was performed
using the mean extinction coefficients for February \citep{frogel98}. 

The narrow band [Fe\,{\sc ii}] image was calibrated by multiplying each filter
transmission curve by the black body spectrum corresponding to the
associated standard star spectral type and then using this integrated
flux as the flux of the standard star in the filter. After
calibration, the emission line image was created by subtracting the
aligned continuum image (Filter 1.06$\mu$m) from the continuum plus emission line image.

\section{Results}

In this section we discuss the morphology of the  narrow-band
[Fe\,{\sc ii}] image and the color map $J-K$ constructed from the
broad band images, comparing our data with narrow-band optical
[O\,{\sc iii}]  and radio images available from previous
studies.  We also present the 2D maps obtained from the IFU
spectroscopy: emission-line distributions and ratios, velocity field and
velocity dispersion maps.

\subsection{Images}

\begin{figure*}
\includegraphics[angle=-90,scale=0.70]{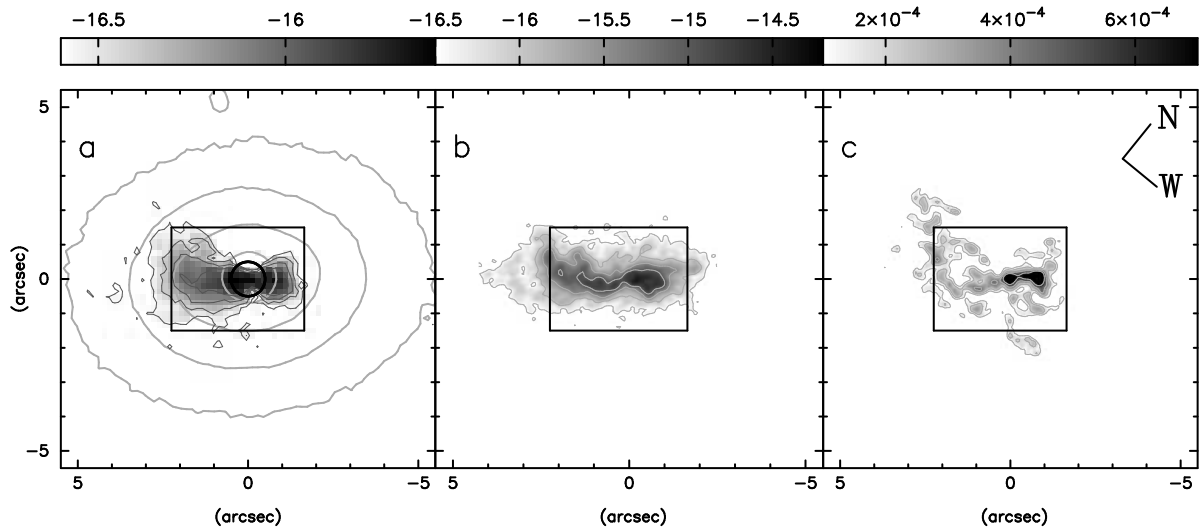}
\caption{Comparison between infrared, optical line emission and radio
continuum maps. The solid straight line is the PA of the [Fe\,{\sc
ii}] maximum extent.  (a) [Fe\,{\sc ii}] (thin contours and gray
scale) overlaid with  the infrared {\it J} band image (thick gray
contours). The flux is given in log(erg~s$^{-1}$~cm$^{-2}$~pix$^{-2})$
according to the gray scale bar at the top.  (b) [O {\sc iii}]
contours and gray scale in arbitrary units. (c) Radio 2-cm contours
and gray scale in arbitrary units. The  black circle indicates the region affected by seeing and the box represent the GNIRS IFU
field of view. The images were rotated to coincide with the orientation of the IFU.} 
\label{painel-FeII-e428}
\end{figure*}

{\em Narrow-band [Fe\,{\sevensize\it II}] image}: In panel {\texttt a}
of Fig.\,\ref{painel-FeII-e428} we present our [Fe\,{\sc ii}] image
which shows emission extended by $\approx$ 3\arcsec\ to the SE and
1\farcs5 to NW along PA$\approx 129\degr$. For comparison, we show
also in Fig.\,\ref{painel-FeII-e428} an {\em HST} narrow band
[O\,{\sc iii}] image (panel {\texttt b}) as well as a radio 2-cm image
(panel {\texttt c}), both from \citet{falcke96}. The [Fe\,{\sc ii}]
image shows a double ``V''-shaped morphology, possibly indicative of a
bi-cone. There is also morphological correlation between [Fe\,{\sc
ii}] and both the [O\,{\sc iii}] and radio 2-cm emission. \citet{falcke96}
present, in addition, an H$\alpha$ map which shows a similar
morphology. The integrated [Fe\,{\sc ii}] flux within a 5\arcsec\
radius is $4.43\times10^{-14}$ erg~s$^{-1}$~cm$^{-2}$.

\begin{figure}
\includegraphics[angle=-90,scale=0.85]{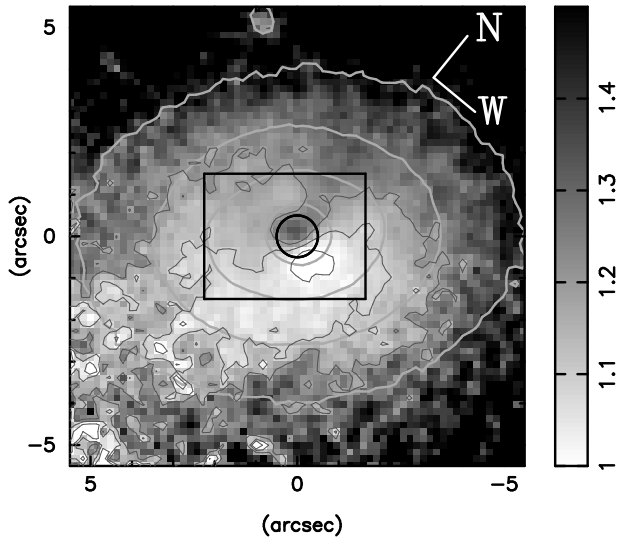}
\caption{$J-K$ colour image (gray scale and thin contours). The
relation between $J-K$ colour and shade of gray is given by the bar at
the top. The {\it J}-band contours are shown overlaid in thick gray lines. The  black circle indicates the region affected by seeing and the box represent the GNIRS IFU
field of view. The images were rotated to coincide with the orientation of the IFU.}
\label{painel-mapa-de-cor-e428}
\end{figure}

\noindent{\em Colour map (J-K)}: The $J-K$ colour map of the inner
10\arcsec~$\times$~10\arcsec\ region of the galaxy is shown in
Fig.\,\ref{painel-mapa-de-cor-e428}. The main feature is a colour
gradient from $J-K = 0.95$ to the SW to $J-K = 1.34$ to the
NE. $J-K\approx 1$ is the color of a late-type stellar population
characteristic of galaxy bulges. Assuming that the redder color is 
due to obscuration by dust, we estimate an average reddening 
$A_V\approx 1.2$ mag for the NE side.
This result also suggests that the NE is the near side of the galaxy,
where dust in the galaxy disk obscures the light from the bulge. In
addition, there is a red feature extending from the nucleus to the N,
possibly due to a dust lane. We estimate $A_V\approx 1.8$ mag for this
feature. The orientation of the  photometric major axis obtained from the J-band image is 129\,${^\circ}\pm$\,2. 

\subsection{2D intensity maps}

\begin{figure*}
\includegraphics[angle=-90,scale=1.43]{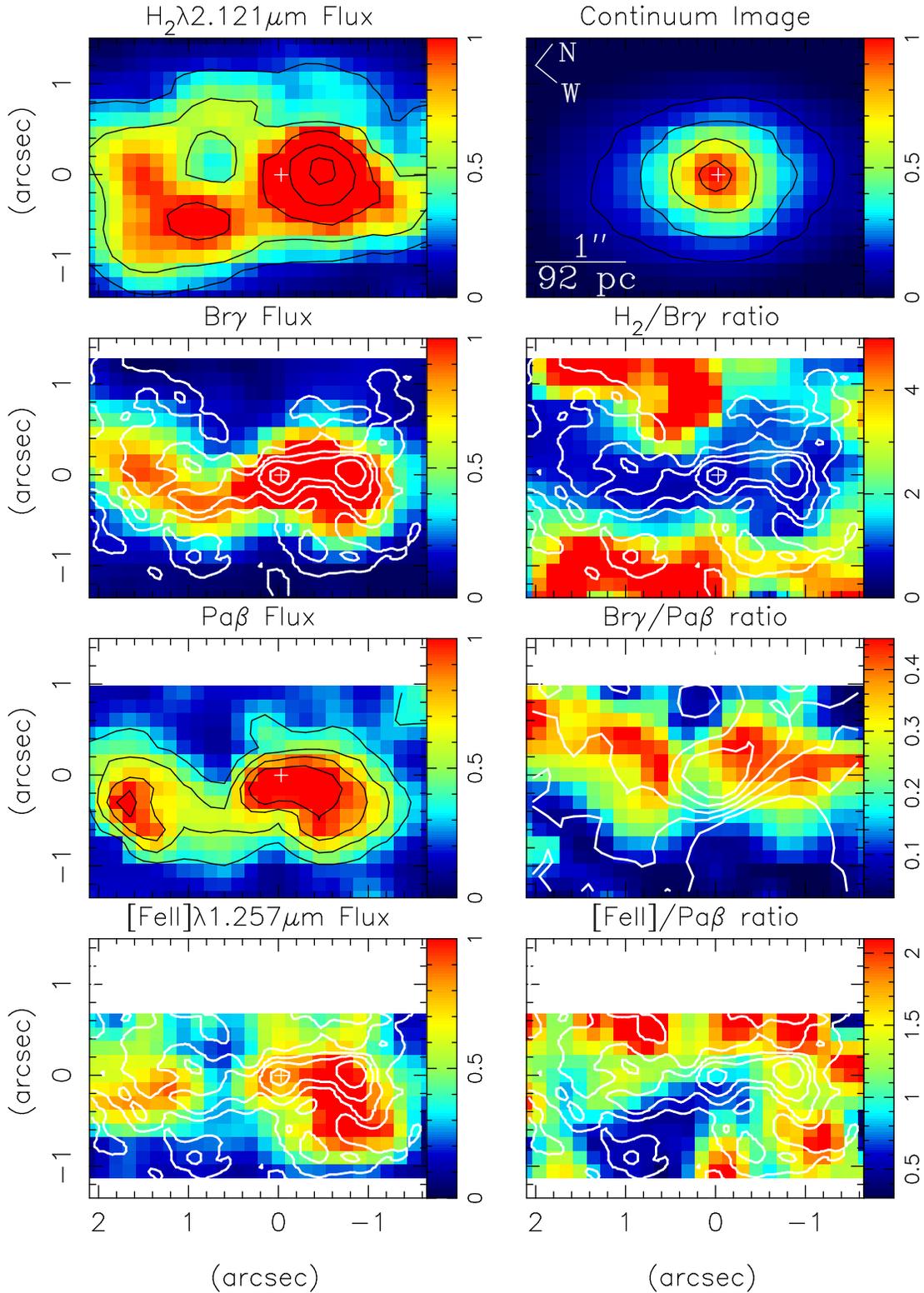}
\caption{Left panels show, from top to bottom, the intensity maps for
H$_2\lambda\,2.121\mu$m, Br$\gamma$,  Pa$\beta$ and [Fe\,{\sc ii}]$\lambda\,1.257\mu$m emission lines with mean uncertainties of 5, 12, 15 and 17\%, respectively. The right panels show, from top to bottom, the reconstructed image for the continuum emission at 2.12$\mu$m, H$_2\lambda\,2.121$/Br$\gamma$
line ratio map, Br$\gamma$/Pa$\beta$ ratio map and [Fe\,{\sc ii}]$\lambda\,1.257$/Pa$\beta$ ratio map.    The thin black contours overlaid  on the H$_2$ and Pa$\beta$  intensity maps and on the continuum image are  isointensity contours. The thick white lines overlaid on the Br$\gamma$,  [Fe\,{\sc ii}], H$_2$/Br$\gamma$ and  [Fe\,{\sc ii}]/Pa$\beta$ maps are the contours of the 2\,cm radio continuum emission from \citet{falcke98} while on the Br$\gamma$/Pa$\beta$ map the contours are from the $J$-$K$\,color map of Fig.\,\ref{painel-mapa-de-cor-e428}. The alignment uncertainty between the radio emission and our 2D maps is $\sim$0$\farcs$2. The spatial orientation and scale are the same for all figures and the cross marks the nucleus (the peak of the continuum emission).}
\label{flux}
\end{figure*}

The final IFU data cube contains 520 spectra, each covering
0$\farcs$15$\times$ 0$\farcs$15, corresponding to 14\,pc$^2$ at the
galaxy. The total IFU field of 3$\farcs$9$\times$3$\farcs$0 thus
corresponds to a region of dimensions 360\,pc$\times$275\,pc at the galaxy. We fitted Gaussian
profiles to the [Fe\,{\sc ii}]$\lambda\,1.257$, H$_2\lambda\,2.121$,
Pa$\beta$ and Br$\gamma$ emission lines in order to obtain the
integrated fluxes, radial velocities (from the peak wavelengths of the
lines) and velocity dispersions (from the widths of the lines). A
$\lambda\,2.12\mu$m continuum image was obtained from the
interpolation of the continuum under the H$_2\lambda\,2.121$ emission
line.

In Fig.\,\ref{flux} we present the 2D maps of the near-IR continuum 
flux and emission line intensities in arbitrary units, with mean
uncertainties of 5, 12, 15 and 17\% for H$_2\lambda\,2.121$,
Br$\gamma$, Pa$\beta$ and [Fe\,{\sc ii}]$\lambda\,1.257$, respectively. In the
same figure we also present the H$_2$/Br$\gamma$, Br$\gamma$/Pa$\beta$ and [Fe\,{\sc
ii}]/Pa$\beta$ line ratios. The nucleus in these maps is defined as
the peak of the continuum light distribution and is identified by the
white cross in each panel of Fig.\,\ref{flux}. In order to investigate
the relation between the radio and near-IR emission, we have overlaid
contours of the radio 2\,cm continuum emission from \citet{falcke98}
on the Br$\gamma$ and [Fe\,{\sc ii}] intensity maps as well as on
the line ratio maps. In order to align
the radio image with our images, we have adopted as the nucleus
in the radio map the peak just to the left of the hot spot, as in Fig.
\ref{painel-FeII-e428}, which we then registered to the peak
of our continuum image. This was also done by \citet{falcke96} for
registering their optical and radio images. They claim that 
the uncertainty between the relative position of the nucleus 
in the optical and radio bands is $\approx$0$\farcs$2.
In our case, the uncertainty may be smaller, due to the
smaller effect of reddening in the near-IR relative to the optical, but
0$\farcs$2 corresponds to only one pixel of our data, thus we may consider such small difference as a coincidence when comparing our images with the radio images. 

The H$_2$ intensity could be measured over most of the IFU field,
while there was not enough signal to measure the Br$\gamma$ and
Pa$\beta$ in a few lines close to the top and bottom edges (~1$\farcs$5 from the
nucleus). The [Fe\,{\sc ii}] emission-line could only be measured over
the inner 1$\farcs$9 along the minor axis of the IFU field, and thus
the corresponding map covers a smaller region than those in the other
emission lines.  All  emission lines are most extended along PA$\approx 129^\circ$,
which is the orientation of the radio jet (and galaxy line of nodes), 
and present a bipolar structure extended to both sides of the nucleus, 
in good agreement with our narrow-band [Fe\,{\sc ii}] image, 
and with both the [O\,{\sc iii}] and radio continuum images presented in
Fig.\,\ref{painel-FeII-e428}. The overlaid contours of the radio
continuum image on the Br$\gamma$ and [Fe\,{\sc ii}] emission-line
maps show a detailed correspondence between the radio and emission-line
structures: the strongest line-emission is observed to the NW, and approximately coincides
with the strongest emission in radio; the emission-line distribution to
the SE bends to the NE, as also observed in the radio map. The peak
of line emission is nevertheless a bit displaced from the radio emission peak:
the line emission peaks at 0$\farcs$6\,NW while the radio emission peaks at
0$\farcs$8\,NW from the nucleus. 

A comparison between the emission-line maps of the H\,{\sc i} with
that of  [Fe\,{\sc ii}] shows that the latter has the strongest emission
displaced to the NW relative to the H\,{\sc i} emission, tracing the
structure of the radio hot spot which bends to the W.

In the case of the H$_2$ emission map, although the above bipolar structure 
is also observed, there is additional emission at the lowest intensity
levels extended beyond the bipolar structure.

\subsection{Line ratio maps}

In order to investigate the main excitation mechanisms of H$_2$
and [Fe\,{\sc ii}], we have constructed the  line-ratio maps  [Fe\,{\sc
ii}]/Pa$\beta$ and  H$_2$/Br$\gamma$ shown in Fig.\,\ref{flux}. The
uncertainties on these ratios range from 0.2 at the nucleus to 0.7
at the top and bottom borders of the IFU field for the first ratio 
and from 0.1 to  0.9 for the second. The uncertainties are thus quite large
at the top and bottom regions of the line ratio maps. We have
overlaid the radio contours also on the line ratio maps.
Along the radio structure, the ratio H$_2$/Br$\gamma$ is
approximately constant, with a value 0.8\,$\pm$\,0.1. Outside the
radio emission region, the H$_2$/Br$\gamma$ ratio increases, reaching
values of 5.3\,$\pm$\,0.9 at 1$\farcs$2 from the nucleus in the
direction perpedicular to the radio axis. The ratio [Fe\,{\sc
ii}]/Pa$\beta$ is 0.9\,$\pm$\,0.3 at the nucleus, and increase up
to $\ge$1.5 at the borders of the radio structure, except to the
South of the nucleus, where it reaches the lowest 
value of 0.5\,$\pm$\,0.2. From the emission-line maps, it can
be observed that this region presents faint levels of both 
[Fe\,{\sc ii}] and radio emission.

We have constructed also the line ratio map Br$\gamma$/Pa$\beta$,
which can be used as a reddening indicator. As we do not
have a relative calibration between the $J$ and $K$ bands,
this map can only be used to investigate the relative reddening
distribution. The lowest ratios are observed to the SW,
while the highest ratios are observed to the NE, the
transition occurring approximately at the major axis of the
galaxy. This behaviour is the same as the one observed in the
$J-K$ map of Fig.\,\ref{painel-mapa-de-cor-e428}, suggesting
that the line emission is subject to the same extinction as
the continuum, with the NE more reddened than the SW side,
consistent with the NE being the near side of the galaxy.
In order to evidence this, we have overlaid
the $J-K$ contours of Fig.\,\ref{painel-mapa-de-cor-e428}
on the Br$\gamma$/Pa$\beta$ ratio map, showing that the reddest $J-K$ values correspond to the highest Br$\gamma$/Pa$\beta$\,ratios.
\subsection{Gaseous kinematics}

\begin{figure*}
\includegraphics[angle=-90,scale=1.45]{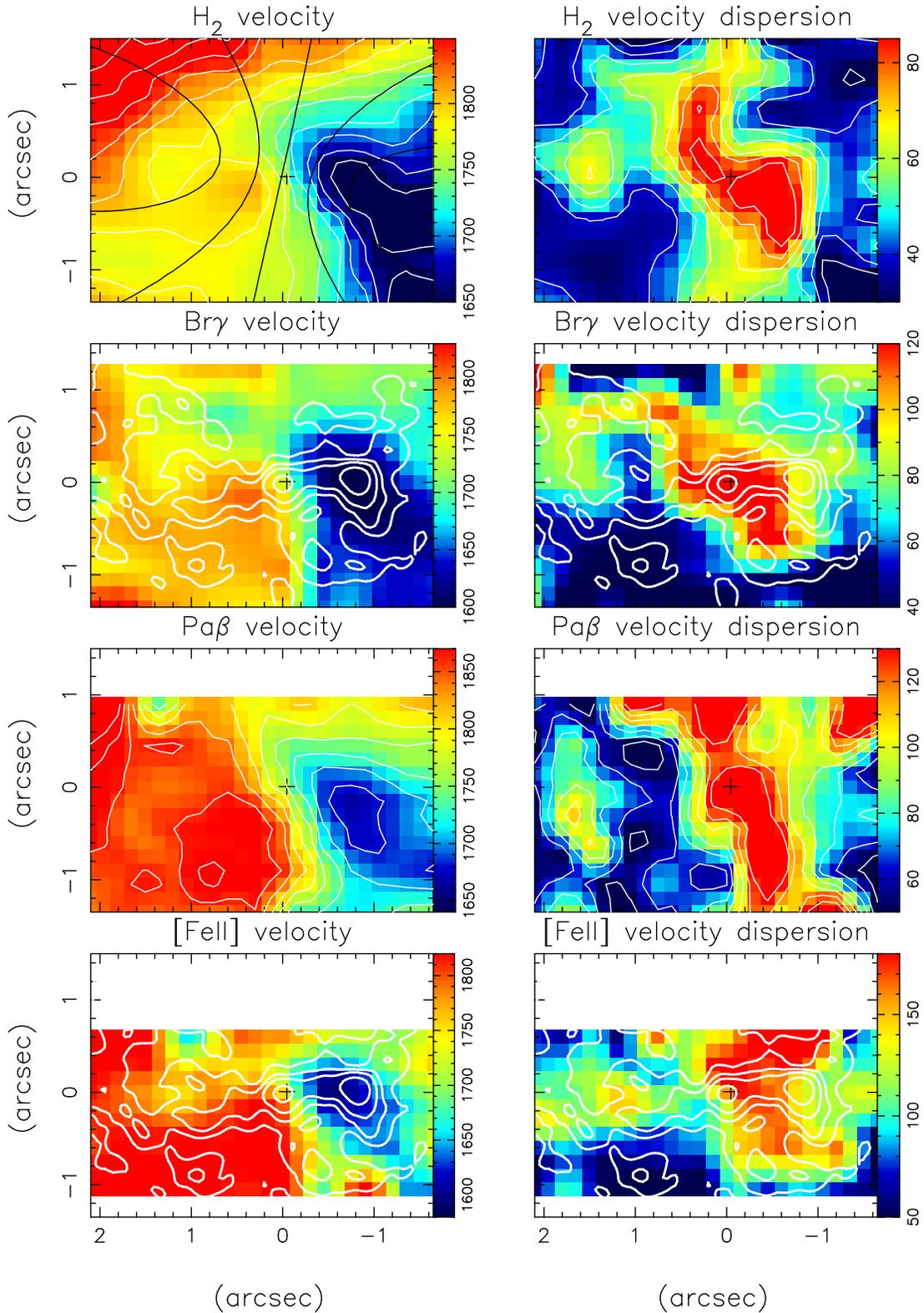}
\caption{Left: From top to bottom, velocity fields of the emitting gas for H$_2$, Br$\gamma$,  Pa$\beta$ and
[Fe\,{\sc ii}]. The mean uncertainty in velocity is less
than 10 km\,s$^{-1}$ for all lines.  Right: Velocity dispersions maps for each emission line with mean uncertainties of 2, 4.5, 14 and 17 km\,s$^{-1}$ for H$_2$, Br$\gamma$,
Pa$\beta$ and [Fe\,{\sc ii}], respectively. The thin white contours on the H$_2$ and Pa$\beta$  maps are isovelocity contours, while the heavy white  contours show overlaid radio emission on the Br$\gamma$ and [Fe\,{\sc ii}] panels. The black lines on the H$_2$ velocity map represent the modeling of the H$_2$ velocity field by a Plummer potential. The spatial orientation and scale are the same as for
Fig.\,\ref{flux}.
} 
\label{vel}
\end{figure*}

Gaseous radial velocity fields were obtained measuring the peak
wavelength of the  emission lines using Gaussian curves to fit 
the emission lines.  The results
are shown in the left column of Fig.\,\ref{vel}, where the red
colors represent redshifts and blue colors represent blueshifts. The
mean uncertainty in the velocities for all lines is less than
10\,km\,s$^{-1}$. An underlying ``rotation pattern'' is present in the
four maps although it is quite clear that in all cases there are other
important kinematic components, evidenced by large deviations from simple rotation. 
We have overlaid the radio contours on the Br$\gamma$ and [Fe\,{\sc ii}] radial velocity
maps, which evidence the strong influence of the radio jet on
the gas kinematics. In particular, the location of the radio hot spot corresponds to the regions of
highest blueshift observed in the gas towards the NW, while there
is also some correspondence between the radio emission and the
redshifts to the SE, mainly observed in the H\,{\sc i} 
and [Fe\,{\sc ii}] velocity maps.

The velocity field most closely resembling circular rotation (the
classical ``spider diagram'') is the one derived from the H$_2$
line. We have tentatively modeled the H$_2$
velocity field as disk rotation in a central Plummer bulge potential
\citep{plummer11}:

\begin{equation}
\Phi=-\frac{GM}{\sqrt{r^2+C_0^2}},
\end{equation}
where $M$ is the mass, $C_0$ the scale length, $r$ is the radius in
the plane of the galaxy and $G$ is the Newton's gravitational
constant. Defining the coordinates of the kinematical center of the system as ($X_0,Y_0$), the
observed radial velocity at the position ($R,\Psi$), where $R$ is the
projected radial distance from the nucleus in the plane of the sky and
$\Psi$ is the corresponding position angle, is  given by 
\[
V_r=V_s+ \]
\begin{equation}
\frac{\sqrt{GM}R\,{\rm sin}\,i\,{\rm cos}^{3/2}\,i}{\lbrace R^2[{\rm sin}^2(\Psi-\Psi_0)+{\rm cos}^2\,i\, {\rm cos}^2(\Psi-\Psi_0)]+C_0^2\,{\rm cos}^2\,i\rbrace^{3/4}},
\end{equation}
where $V_s$ is the systemic velocity, $i$ is the inclination of the
disk ($i=0$ for face on disk) and $\Psi_0$ is the position angle of
the line of nodes \citep{storchi-bergmann96}. The equation above
contains seven free parameters, including the kinematical center, that
can be determined by fitting the model to the observations. This was
done using a non linear least-squares algorithm, in which initial
guesses are given for the free parameters. The isovelocity curves
representing the best model fit to the data are shown as black lines
in  the upper left panel of Fig.\,\ref{vel}, while the white
contours trace the isovelocity curves of the observed velocity
field. It can be seen that circular rotation is not a good
representation of the velocity field, particularly along the radio
axis,  evidentiating the presence of other kinematic components in the gas
besides rotation. The kinematic center obtained from the fit coincides
with the location of the peak of the continuum light distribution
within the uncertainties. 

The parameters derived from this fit are:
the systemic velocity -- 1752\,$\pm$\,2.8\,km\,s$^{-1}$ is
54\,km\,s$^{-1}$ larger than the one listed in NASA/IPAC Extragalactic
Database (NED), while the inclination of the disk --
54.8$^\circ$\,$\pm$\,3.9$^\circ$  is 10$^\circ$ higher than the
photometric one we obtain from the {\it J}-band image ($\rm
cos^{-1}(\frac{b}{a})$, where $a$ and $b$ are the semi-major and
semi-minor axes of the ellipse fitted to the outermost isophote of
Fig.\,\ref{painel-mapa-de-cor-e428}). We obtain the value 1.02$\pm$0.13
$\times$\,10$^9$\,M$_\odot$ for the bulge mass and
108.5\,$\pm$\,12.9\,pc for the bulge scale length.  The orientation of
the line of nodes -- 119.5$^\circ$\,$\pm$\,1.4$^\circ$ is 10$^\circ$
smaller than that of the photometric major axis 

We have also measured the full width at half maximum (FWHM) of the
observed emission lines, from which we obtained the velocity
dispersion as $\sigma=\rm\frac{FWHM}{2.35}$. The corresponding
$\sigma$ maps for each line are shown in the right panels of
Fig.\,\ref{vel}. The mean uncertainties in $\sigma$ values are 2, 4.5, 14 and 17 km\,s$^{-1}$ for H$_2$, Br$\gamma$, Pa$\beta$ and [Fe\,{\sc ii}], respectively.  We have again overlaid
the radio contours on the Br$\gamma$ and [Fe\,{\sc ii}] $\sigma$\,maps. 
It can be observed that the regions with the largest $\sigma$ are displaced
relative to the regions with strongest radio emission: for example,
the largest $\sigma$ values are observed between the radio nucleus
and the hot spot in the Br$\gamma$ $\sigma$\,map. Similarly, in 
the [Fe\,{\sc ii}] $\sigma$\,map the highest values avoid
also the locations of the radio emission knots.

A comparison between the $\sigma$\,maps of the different lines 
show that the H$_2$ velocity dispersion values are lower than those of [Fe\,{\sc
ii}] and H\,{\sc i} lines,  indicating that the gas emitting the two
latter lines is more perturbed than the molecular gas. 

\begin{figure*}
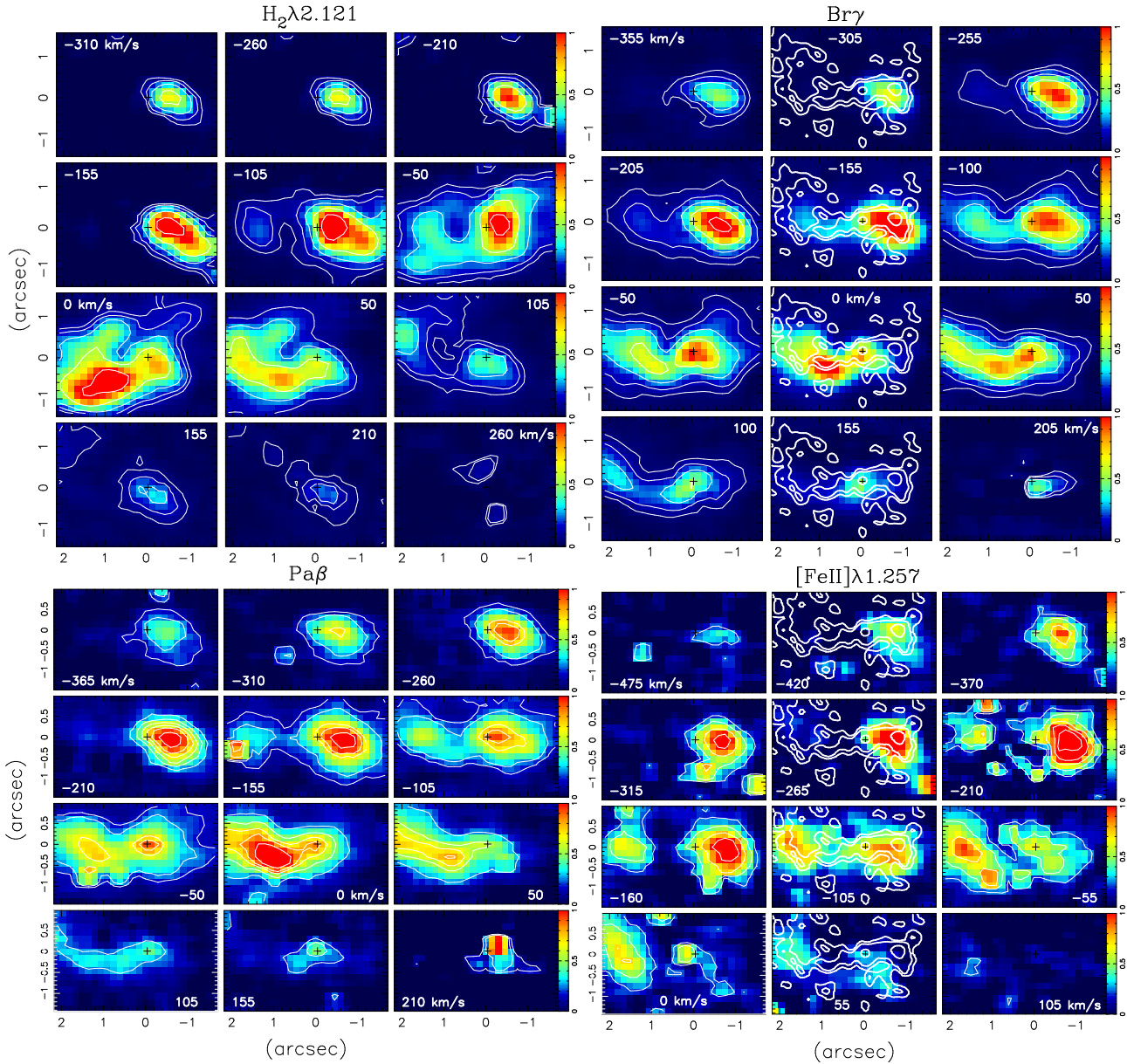

\includegraphics[angle=-90,scale=0.65]{int_velh2.eps}
\includegraphics[angle=-90,scale=0.65]{int_velbr.eps}
\includegraphics[angle=-90,scale=0.63]{int_velpa.eps}
\includegraphics[angle=-90,scale=0.63]{int_velfe.eps}
\caption{Velocity slices across the emission lines in $\sim$\,50km\,s$^{-1}$ velocity bins centered on the velocities indicated in each panel. The black cross
marks the position of the nucleus (peak of continuum emission).  Top:
H$_2$ (left),  Br$\gamma$ (right); Bottom:  Pa$\beta$ (left), 
[Fe\,{\sc ii}] (right). We present flux contours for each slice except  for the middle panels of Br$\gamma$ and [Fe\,{\sc ii}] where we present the radio continuum contours. The spatial orientation and scale are the same as for Fig.\,\ref{flux}.} 
\label{slices}
\end{figure*}

\subsection{Emission-line ``Tomography''}

The relatively high spectral resolution of the data has allowed us to
slice our data cube into a sequence of velocity bins along each
emission line profile, providing a better sampling of the gas
kinematics, not restricted to the peak wavelength of the
emission-line, but including also the wings. The slices were obtained
after the subtraction of the continuum determined as averages of the
fluxes from  both sides of each emission-line. Each slice corresponds to
a velocity bin of $\approx$50\,km\,s$^{-1}$, selected to include two
pixels of the data cube and to correspond approximately to the nominal
spectral resolution of the data. The obtained slices are shown in
Fig.\,\ref{slices}, where each panel presents flux contour levels in
arbitrary units for each velocity slice and each emission line, and
where the nucleus, defined as the peak of continuum emission, is
marked with a cross. 
The zero velocity is adopted as the one
corresponding to the peak wavelength of H$_2$
within an aperture of 3$\times$3\,pixels centered at the nucleus. We have overlaid the contours of the radio image on the central panels the Br$\gamma$ and [Fe\,{\sc ii}] emission line slices in Fig. \,\ref{slices}. 

The velocity distributions are somewhat different for the
different emission lines.
For H\,{\sc i}, as the slices trace gas from negative
(blueshift) to  positive (redshift) residual velocities, the peak in
the flux distribution moves from NW to SE,
approximately following the path traced by the radio emission. The
blueshifts observed in H\,{\sc i}, reach $\approx$\,300\,km\,s$^{-1}$, approximately at the location of the radio hot spot, between 0$\farcs$6 and 0$\farcs$8\,NW of the nucleus. Some blueshifts are also observed to the SE following the radio structure which bends upwards (see Fig. \,\ref{slices}).
Redshifts are mostly observed at the nucleus, where they reach
$\approx$200\,km\,s$^{-1}$ and 
to the SE where they reach smaller values of $\approx$100\,km\,s$^{-1}$ at
1--2$^{\prime\prime}$SE. The redshifted emission to the SE also traces the radio structure. Thus, to the SE, we observe both blueshifts and redshifts following the radio structure. 

For H$_2$, the behaviour is similar to that observed for H\,{\sc i}, only that there is additional emission beyond the radio structure. The highest blueshifts, of up to $\approx$400\,km\,s$^{-1}$ are
observed in the [Fe\,{\sc ii}] emitting gas,
which shows a more bipolar structure in the velocity slices than the other emission lines. Throughout this structure, we observe mostly blueshifts, even at the nucleus and to SE.

In summary, we could say that the [Fe\,{\sc ii}] velocity field shows predominantly a bipolar structure, which is  also observed in the  H$_2$ and H\,{\sc i} for the highest velocity gas. For velocities between -100 and 100 km\,s$^{-1}$, the structure in H\,{\sc i}
is best described as linear,
with both blueshifts and redshifts observed to either side of the
nucleus. In the case of H$_2$, in this low velocity range, the
emission at low intensity levels is spread over most of the IFU field. 

\section{Discussion}
\subsection{Gas kinematics}

\begin{figure}
\centering
\includegraphics[scale=0.45]{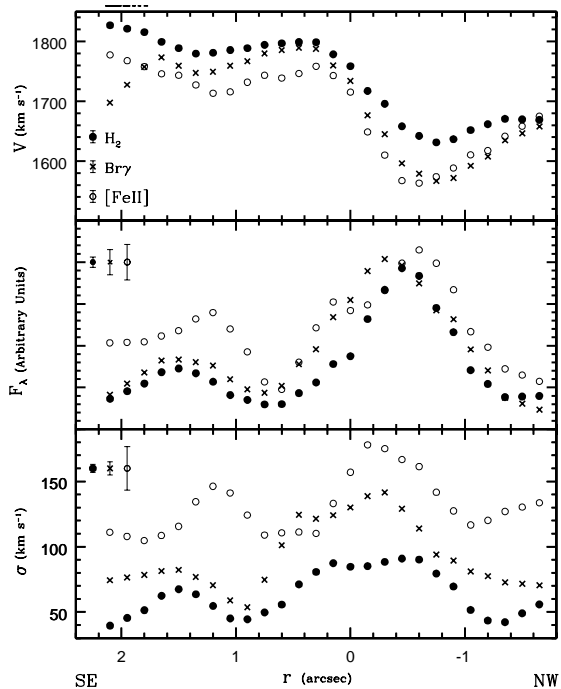}
\caption{ One-dimensional cuts along the radio axis. Top: Radial velocity for H$_2$ Br$\gamma$ and [Fe\,{\sc ii}]. Middle: Emission line fluxes  in arbitrary units. Bottom: Velocity dispersion values. }

\label{radial}  
\end{figure}

Although all the velocity fields show evidence of a rotation pattern,
the blueshifted side shows a more abrupt gradient going from the
systemic velocity at the center to $-$100\,km\,s$^{-1}$ at only  
0$\farcs$6 (55\,pc) from it towards the NW,  while to the SE
equivalent redshift is barely reached at the border of the IFU field
(210\,pc from the nucleus). In the top panel of Fig.\,\ref{radial} we present one-dimensional cuts of the radial velocity field along the radio jet for the H$_2$, Br$\gamma$ and [Fe\,{\sc ii}] emission lines, in which  we can clearly see this behaviour. 

By comparing the line intensity maps (Fig.\,\ref{flux}) with the
velocity fields (Fig.\,\ref{vel}), we verify that the peak
blueshifts approximately coincide with the peak intensities, 
suggesting that the flux enhancement is produced by the 
compression provided by the NW radio jet
(\citealt{falcke98}), which should be at least partially oriented toward
us in order to account for the observed blueshift. A more detailed analysis of the correspondence between the line fluxes and velocities (but along one axis) can be done using one-dimensional cuts shown in Fig.\,\ref{radial}, by comparing the top and middle panels. In this comparison it can be observed that the [Fe\,{\sc ii}] line flux presents a peak at the location of the strongest blueshift, between 0$\farcs$6 and 0$\farcs$8 NW of the nucleus, another smaller peak at the nucleus and another one at 1\farcs2 SE of the nucleus, at a location coincident with a ``residual blueshift" -- a decrease in the velocity observed in the redshifted side in the top panel of Fig.\,\ref{radial}. The fluxes of the H$_2$ and Br\,$\gamma$ emission lines also show peaks at $\approx$0$\farcs$4\,NW, thus somewhat closer to the nucleus than the locations of strongest blueshifts and between 1 and 2$^{\prime\prime}$ SE of the nucleus, approximately at the locations of the residual blueshift  described above, also observed in the H$_2$ and Br\,$\gamma$ emission lines. Similar displacements between the radio and optical emission have been  observed by \citet{falcke98} in their comparison between HST [O\,{\sc iii}]$\lambda$5007 and [N\,{\sc ii}s]+H$\alpha$ emission-line images and the radio images of ESO\,428-G14: they found that the emission-line fluxes are enhanced, in the form of knots and filaments in structures shaped as cocoons surrounding the radio jets.  This close association between emission line distributions and radio structures is nevertheless  not always observed  leading some authors to conclude that there is no connection between the kinematics of the NLR and radio jets \citep{kaiser00,das05,das06}.

In the case of ESO\,428-G14 the good correspondence between the radio and emission line structures, also observed in the velocity slices of Fig. \ref{slices}
indicates that the radio jet has an influence on all the emitting gas, although the somewhat  distinct kinematics observed in  the different emission lines suggest varying contribution of the radio jet to the line emission. The [Fe\,{\sc ii}] line kinematics is
dominated by two outflowing structures towards the end of  the radio jets; the H\,{\sc i}
kinematics has contribution from all emitting regions along the radio jet; and the
H$_2$ line presents the less bipolarity, more closely resembling a
rotation pattern. We interpret these differences as due to a larger
disk rotation component contribution to kinematics of the H$_2$ line,
an increasing contribution of the radio jet perturbations to the
kinematics of the H\,{\sc i} line emission,  and a even larger contribution
from the radio jet to the kinematics of the [Fe\,{\sc ii}] line.

The influence of the radio jet on the emitting gas can also be observed in the velocity dispersion ($\sigma$) maps (Fig.\,\ref{vel}) and in the one-dimensional cuts along the radio axis shown in the lower panel of Fig.\ref{radial}. Increase in the $\sigma$ values are observed from  0$\farcs$4\,SE to 0$\farcs$6\,NW along the radio axis and between 0$\farcs$8 and 1$\farcs$8 to SE of the nucleus. At least to the NW, the largest $\sigma$ values are thus observed between the nucleus and the radio hot spot, and do not coincide with the emission line flux and velocity peaks, but is shifted towards the nucleus. 
To the SE, the highest $\sigma$ values seem to approximately coincide with the peaks in the line intensity and velocity (the residual blueshifts relative to the redshifts of the gas to SE). We interpret this result as
due to the interaction of the radio jet with ambient gas which
produces  a broadening of the emission lines due the momentum
transfered to the gas by the jet.
The different behaviour of the $\sigma$ values to the NW and to the SE can be understood as due to  a shock between the radio jet and a dense ISM to the NW which blocks the
radio jet  and stirs the gas not only at the shock front but also 
behind the shock, which is observed as a larger velocity dispersion.  The presence of a shock front is evidenced by the the abrupt termination of the radio jet at the hot spot and the enhanced line emission at this location. 
The interaction of the radio jet with the ISM is milder to the SE, where the radio
jet stirs the gas but is not blocked penetrating farther into the ISM.

It can also be noted that the overall velocity dispersion values are
lowest for H$_2$ (30 to 85\,km\,s$^{-1}$), and highest for the
[Fe\,{\sc ii}] (50 to 180\,km\,s$^{-1}$), indicating that the
[Fe\,{\sc ii}] traces the most disturbed gas, while H$_2$ traces the
less disturbed gas. This result is also consistent with those obtained
from the velocity fields discussed above, supporting a different origin for the gas emitting the different lines. Distinct kinematics for H\,{\sc i}, H$_2$ and  [Fe\,{\sc ii}] has been observed also in previous near-IR studies  of other AGNs (e.g. \citealt{wilman00}, \citealt{storchi-bergmann99}), where  the broadest [Fe\,{\sc ii}] emission has been attributed to gas cloud interactions with the radio jet.

\subsection{H$_2$  emission origin}

The H$_2$ lines can be excited in two different ways: fluorescent
excitation through absorption of soft-UV photons (912--1108 \AA) in
the Lyman and Werner bands \citep{black87} and collisional excitation,
namely inelastic collisions between molecules in a warm ($T \ga 1000$
K) gas. The heating necessary to allow collisional excitation may be
provided by: shocks \citep{hollenbach89}, 
X-rays \citep{maloney96} or dense photodissociation regions
by UV photons. The method commonly used to differentiate collisional
excitation from fluorescence is based on the flux ratios of H$_2$
lines in the {\it K} band. Based on measurements of such ratios for a
large sample of galaxies,  \citet{rodriguez-ardila05} concluded that
fluorescence is not important for AGNs supporting previous studies by
\citet{veilleux97}, \citet{moorwood90} and \citet{fischer87}.  

\citet{quillen99}, using {\em HST}, imaged a sample of 10 Seyfert
galaxies in H$_2$ and detected extended emission in six of them. For
three galaxies they found H$_2$ emission in the extended narrow line
region, coincident with [O\,{\sc iii}] and H$\alpha +$\,[N\,{\sc ii}]
line emission. For these galaxies they discarded UV excitation as the
dominant excitation process on the basis of the H$\alpha$ to H$_2$
ratio, and argued that slow shocks are the most likely process to
explain these line ratios. They compiled spectroscopic observations of
molecular hydrogen from \citet{koornneef96}, \citet{ruiz97} and
\citet{veilleux97} and looked for correlations with radio 6-cm and
hard X-ray flux. They found no correlation with X-rays, suggesting
this is not the dominant H$_2$ excitation mechanism, and found a weak
correlation with radio 6-cm, suggesting that no single mechanism is
likely to be responsible for the molecular hydrogen excitation in
Seyfert galaxies. 

In a paper aimed at explaining the strong H$_2$ emission of NGC\,6240, \citet{draine90} argued that shock speeds higher than the H$_2$ $\sigma$ values of ESO\,428-G14 (50$\le \sigma\le$80\,kms$^{-1}$) dissociates the H$_2$ molecule.  We thus conclude that faster shocks, although present, as indicated by the larger [Fe\,{\sc ii}] $\sigma$ values, destroy the  H$_2$ molecule, and this is why we do not see broader H$_2$ emission. 
\citet{draine90} propose that most of the H$_2$ line emission originates in molecular gas with is heated by transient X-ray irradiation. Hard X-rays from the AGN have been also proposed by \citet{wilman00} and \citet{bellamy04} as the dominant excitation mechanism of H$_2$ emission in Cygnus\,A. 

Our H$_2$ flux distribution and kinematics suggest some association
with the radio morphology, in particular at the radio hot spot, but
there is also additional emission spread throughout the IFU field. The
emission line ratio H$_2\lambda 2.121$/Br$\gamma$ is $\approx$\,0.8
along the radio jet increasing outwards, in the direction
perpendicular to the radio jet, to $\approx$5. In Starbursts, where
the main heating agent is UV radiation, H$_2\lambda 2.121$/Br$\gamma <
0.6$,  while for AGNs this ratio is larger (0.6 $<$ H$_2\lambda 2.121
$/Br$\gamma  <$ 2) because of the additional H$_2$ excited  by shocks
or by X-rays from the AGN
\citep{rodriguez-ardila05,storchi-bergmann99}. The value
$\approx$\,0.8 observed for ESO\,428-G14 along the radio jet is thus
typical of AGNs. 

The higher H$_2\lambda 2.121$/Br$\gamma$ line ratio values  observed outside the jet could be due to some X-rays from the AGN still reaching these regions while the UV-photons needed to excite the H\,{\sc i} are in much less number producing only very faint Br$\gamma$ emission. 

From the discussion above we conclude that the collisional excitation necessary to  
produce the H$_2$ emission may be provided by the interaction of the radio jet with the 
circumnuclear interstellar medium (hereafter ISM) or by heating produced by the X-rays 
emitted by the AGN. The enhanced H$_2$ $\sigma$ values in the hot spot region reveal that the jet is depositing kinetic energy on the gas.  This energy can be estimated as follows. The instrumental broadening is 21 km\,s$^{-1}$. Subtracting this in quadrature, we
obtain a minimum $\sigma$ for H$_2$ of 34 km\,s$^{-1}$ (away from the radio jet), and a maximum value at
the hot spot region of 77 km\,s$^{-1}$ . (Uncertainties in these values are $\sim$\,10
km\,s$^{-1}$.)  Velocities of the order of the latter provide enough kinetic energy to excite the H$_2$ molecule.  Assuming that the H$_2$ emitting regions with smallest 
$\sigma$ (blue region in the top right  panel of  Fig. \ref{vel}) are solely excited by X-rays and that the X-ray contribution is the same as in the jet region, we can estimate the increase in kinetic energy provided by the radio jet  as $\rm(77/34)^2=5$. 
 Under the above assumptions, one could conclude that $ \sim$83\% (5/6) of the H$_2$ excitation  is due to shocks at the hot spot regions (red regions in the top panel of
Fig. \ref{vel}).  In the yellow regions SE of the nucleus, and the regions surrounding the radio jet region, the kinetic energy ratio is 2.7, thus the jet excitation still dominates ($\sim$73\%) there.  But these 
are only upper limits because the X-ray flux in the jet region may be larger. In a 
``photoionization scenario" by a ``clumpy" torus atmosphere \citep[eg.][]{evans93,konigl94}, the ionization state of the gas would be larger along the collimation axis of 
the  torus and would drop with increasing distance from this axis. This could explain why 
the H$_2$ emission is distributed over a wider area, in gas which is irradiated by an 
attenuated continuum.

\subsection{[Fe\,{\sc ii}] emission origin}

A number of papers have presented reviews and detailed studies of the
physical conditions of the [Fe\,{\sc ii}] emitting gas in starburst
and AGN galaxies, such as \citet{mouri90}, \citet{mouri93}
and \citet{rodriguez-ardila05}. In these studies it is
argued that the [Fe\,{\sc ii}]/Br$\gamma$ ratio  is controlled by the
ratio between the volumes of partially to fully ionized gas regions,
as [Fe\,{\sc ii}] emission is excited in partially ionized gas. Such
zones in AGN are created by power law photoionization (including X-rays emitted by the AGN) or shock 
heating. These two processes are discriminated by the electron
temperature of the  [Fe\,{\sc ii}] region: $T_e\approx\rm8000\,K$ for
photoionization and $T_e\approx\rm6000\,K$ for shocks. A contribution
from shocks produced by radio jets is supported by results such as
those of \citet{forbes93} which have found a correlation between the
[Fe\,{\sc ii}] and 6-cm radio emission in radio galaxies. Nevertheless
\citet{simpson96} have argued that the dominant excitation mechanism
of [Fe\,{\sc ii}] is photoionization with shock excitation accounting
for only $\approx$ 20\,\% of the [Fe\,{\sc ii}] emission in AGN. 

The nature of the excitation mechanism can be investigated using  emission line ratios. \citet{rodriguez-ardila05} show that [Fe\,{\sc ii}]/Pa$\beta$ is
smaller than 0.6 for Starburst galaxies and higher than 2 for
supernova remnants, for which shocks are the main excitation
mechanism. The Seyfert galaxies have values in between 0.6 and 2
suggesting that this ratio can be used as a measure of the relative
contribution of photoionization and shocks. For ESO\,428-G14
1$\pm$0.27$<$[Fe\,{\sc ii}]/Pa$\beta<$2$\pm$0.73 in the region
co-spatial   with the radio jet,  indicating a range of relative
contribution of photoionization and shock excitation for the [Fe\,{\sc
ii}] line, with regions where this ratio is $\approx$\,2 been
dominated by shocks. In the region corresponding to the radio hot spot
this ratio ranges from 1.5 to 2 supporting a large contribution from
shocks. 

In order to estimate an upper limit  on the contribution of the radio jet in the excitation of the [Fe\,{\sc ii}], we use the same reasoning as in the case of H$_2$ emission above, assuming that the lowest $\sigma$
observed for this line (dark blue regions in the bottom right panel of Fig.\,\ref{vel})  is produced by  X-ray excitation. We obtain a ratio
between maximum and minimum kinetic energies of $\rm(180/60)^2=9$, thus 90\%
of [Fe{\sc ii}] emitted in the hot spot region (red region in bottom right
panel of Fig.\,\ref{vel}) is excited by the radio jet under the above
assumptions. In other regions of the jet (yellow-green regions) where $\sigma\sim\rm130\,km\,s^{-1}$, the
percentage of gas emission due to radio jet excitation falls to 80\%. But these  are also upper limits because as pointed above for H$_2$, the X-ray emission may be stronger along the radio jet,  providing a larger contribution for the [Fe{\sc ii}] emission excitation  than in the regions with the lowest $\sigma$ values.
On the other hand, the stronger association of the [Fe{\sc ii}] flux and kinematics with the 
radio structure supports a larger contribution of the radio jet to the [Fe{\sc ii}] excitation 
relative to that of H$_2$.

\section{Summary and Conclusions}

We have analyzed near-IR {\it J} and {\it K} band  2D spectra of the
inner $\approx$300\,pc of the Seyfert galaxy ESO\,428-G14 obtained
with the Gemini GNIRS IFU. The fine angular sampling of this
instrument and the high spectral resolution of the observations
provided a spatial sampling of 14\,pc$^2$ at the galaxy and a velocity
resolution of $\approx$20\,km\,s$^{-1}$. These characteristics and 2D
coverage has allowed us to obtain gas flux distributions and ratios as
well as to map the gas kinematics in the four strong
emission lines [Fe\,{\sc ii}]$\lambda 1.257 \mu$m, Pa$\beta$, H$_2 \lambda 2.121 \mu$m and Br$\gamma$. Such 2D mapping at the spatial and spectral resolution of
our observations and in the near-IR is unique in the literature so
far.

We have used the spectroscopic data in conjunction with a [Fe\,{\sc
ii}]$\lambda 1.257 \mu$m narrow-band image and $J-K$ color map
obtained with OSIRIS at the CTIO Blanco Telescope. The [Fe\,{\sc ii}]
narrow-band image shows a bi-polar structure which correlates well
with the structure observed in a previous published radio map, as well
as with that observed in a narrow-band [O\,{\sc iii}] HST image. This
bi-polar structure is oriented approximately along the photometric
major axis of the galaxy at PA$\approx$129$^\circ$. Under the
assumption that redder color is due to dust screening, we conclude
from the $J-K$ color map that NE is the near side of the galaxy. 

The IFU observations show that not only the [Fe\,{\sc ii}] flux
distribution, but also Pa$\beta$ and Br$\gamma$ correlate with the
radio emission, all showing elongated morphologies along the radio
axis. The H$_2$ emission has an additional component more
spread over the disk of the host galaxy. 

The gas kinematics shows a pattern which, at first sight, seems to be
due to rotation in a disk, but which is clearly disturbed by
non-circular motions. The analysis of the gas kinematics is
complicated by the fact that the galaxy major axis is approximately
aligned with the radio axis. The radio emission morphology suggests
that a radio jet is the cause of outflows observed in velocity
``slices" obtained along the gas emission line profiles: blueshifts of
$\sim$ 300--400\,km\,s$^{-1}$ are observed in emission associated
with a radio hot spot at $\sim$0\farcs8 NW of the nucleus, while mostly
redshifts and some blueshifts observed to the opposite side 
of the nucleus are associated to fainter radio emission observed up to 
2$^{\prime\prime}$ SE. From this velocity pattern, 
as well as from the emission-line ratios we
conclude that the radio jet leaves the nucleus at a small angle relative to the plane of the galaxy, with the NW side slightly 
directed toward us and the counter-jet directed
away from us, being partially hidden by the disk of the galaxy. Some redshift observed to the NW and some blueshift observed to the SE shows that the line emission is observed to both sides of the plane of the galaxy, confirming
that the angle between the radio jet and the plane of the galaxy is small and emission associated with the jets occurs on both sides of the galaxy plane.
The reddening distribution obtained from the Br$\gamma$/Pa$\beta$ emission-line ratio coincides with that obtained from the continuum, confirming that the line emission comes from circumnuclear gas located essentially in the galactic plane.  

The velocity dispersion maps show the highest values between the nucleus and the regions of strongest radio emission (the radio hot spots). We interpret this result as due to kinetic energy deposited by ratio jet in the circumnuclear ISM, producing a compression in the gas and emission enhancement just beyond this compressed region. 

While the [Fe\,{\sc ii}] emission is dominated by the highest
velocities at the two hot spots in opposite sides of the nucleus, the
H\,{\sc i} emission has an important contribution also from lower
velocities, observed all along the radio emission structure. The H$_2$
emission has significant emission contribution at low velocities
spread throughout the observed region, apparently coming from the disk
of the host galaxy.

From the observed kinematics, we conclude that the radio jet has a fundamental role in shaping the emission line region as it interacts with the galaxy ISM surrounding the galaxy nucleus. 
We have used the 2D velocity dispersion maps  to estimate the kinetic energy deposited in the circumnuclear ISM by the radio jet relative to regions away from the jet. Assuming that the H$_2$ and [Fe\,{\sc ii}] excitation  in the latter regions is dominated by X-rays, and that the X-rays excitation is the same in the jet region, we obtain contributions by shocks of up to 
 80\,-\,90\% for the [Fe\,{\sc ii}] and up to 70\,-\,80\% for H$_2$ in the jet region. These are however, upper limits due to the fact that the X-rays contribution along the jet axis may be larger than away from it. The stronger association of the [Fe\,{\sc ii}] emission and kinematics with the  radio structure supports a larger contribution of the radio jet to the excitation of [Fe\,{\sc ii}] than to  that of H$_2$.


\section*{Acknowledgments}
We thank an anonymous referee for useful comments and suggestions which helped to 
improve the paper.
We acknowledge valuable discussions with A. S. Wilson and
H. R. Schmitt and thank H. Falcke for sending the radio image.
 Based on observations obtained at the Gemini
Observatory, which is operated by the Association of Universities for
Research in Astronomy, Inc., under a cooperative agreement with the
NSF on behalf of the Gemini partnership: the National Science
Foundation (United States), the Particle Physics and Astronomy
Research Council (United Kingdom), the National Research Council
(Canada), CONICYT (Chile), the Australian Research Council
(Australia), CNPq (Brazil) and CONICET (Argentina). This research has
made use of the NASA/sIPAC Extragalactic Database (NED) which is
operated by the Jet Propulsion Laboratory, California Institute of
Technology, under contract with the National Aeronautics and Space
Administration.

\label{lastpage}

\end{document}